# Commentary Article

# Towards a Framework for the Design, Implementation and Reporting of Methodology Scoping Reviews

Glen P. Martin [1*]; David A. Jenkins [1,2]; Lucy Bull [3,4]; Rose Sisk [1]; Lijing Lin [1]; William Hulme [1]; Anthony Wilson [5]; Wenjuan Wang [6]; Michael Barrowman [1]; Camilla Sammut-Powell [1]; Alexander Pate [1]; Matthew Sperrin [1]; Niels Peek [1,2]; On Behalf of The Predictive Healthcare Analytics Group

1. Division of Informatics, Imaging and Data Science, Faculty of Biology, Medicine and Health, University of Manchester, Manchester Academic Health Science Centre, Manchester, UK

2. NIHR Greater Manchester Patient Safety Translational Research Centre, University of Manchester, Manchester, UK

3. Manchester Epidemiology Centre versus Arthritis, Centre for Musculoskeletal Research, Manchester Academic Health Science Centre, University of Manchester, Manchester, UK

4. Centre for Biostatistics, Manchester Academic Health Science Centre, University of Manchester, Manchester, UK

5. Adult Critical Care, Manchester University Hospitals NHS Foundation Trust, Manchester UK.

6. Department of Population Health Sciences, Faculty of Life Science and Medicine, King's College London, UK

**Word Count:** 2975

**Corresponding Author**
Dr Glen Philip Martin
Lecturer in Health Data Science
Division of Informatics, Imaging and Data Science, Faculty of Biology, Medicine and Health, University of Manchester,
Vaughan House, Manchester, M13 9GB, United Kingdom
Email: glen.martin@manchester.ac.uk

# Abstract

**Background**

In view of the growth of published papers, there is an increasing need for studies that summarise scientific research. An increasingly common review is a "Methodology scoping review", which provides a summary of existing analytical methods, techniques and software, proposed or applied in research articles, which address an analytical problem or further an analytical approach. However, guidelines for their design, implementation and reporting are limited.

**Methods**

Drawing on the experiences of the authors, which were consolidated through a series of face-to-face workshops, we summarise the challenges inherent in conducting a methodology scoping review and offer suggestions of best practice to promote future guideline development.

**Results**

We identified three challenges of conducting a methodology scoping review. First, identification of search terms; one cannot usually define the search terms *a priori* and the language used for a particular method can vary across the literature. Second, the scope of the review requires careful consideration since new methodology is often not described (in full) within abstracts. Third, many new methods are motivated by a specific clinical question, where the methodology may only be documented in supplementary materials. We formulated several recommendations that build upon existing review guidelines. These recommendations ranged from an iterative approach to defining search terms through to screening and data extraction processes.

**Conclusion**

Although methodology scoping reviews are an important aspect of research, there is currently a lack of guidelines to standardise their design, implementation and reporting. We recommend a wider discussion on this topic.

**Key Messages**

- Reviews that aim to summarise existing analytical methods, techniques or software, proposed or applied in research articles, which address an analytical problem or further an analytical approach, are becoming an essential component of research.
- Guidelines for the design, implementation and reporting of such "methodology scoping reviews" are limited.
- By collating experiences of the authors, we here present several recommendations for conducting methodology scoping reviews, which build upon existing review guidelines where possible.

**Background**

The scientific literature is growing at a large and increasing rate, and the drive towards evidence-based practice increasingly relies on synthesising evidence within a given research field. In quantitative research, analytical methods are used to learn from data, and through the full translational pipeline eventually turn this understanding into evidence about a particular situation that supports future decisions. Here, systematic reviews are often regarded as the gold standard by which such evidence can be summarised and/or synthesised. The position of systematic reviews atop the 'hierarchy of research evidence' [1] ensures that there are well-established guidelines for designing, conducting and reporting systematic reviews (e.g. PRISMA [2], amongst many others), driven by groups such as the Cochrane Collaboration.

Increasingly, there is a need to summarise the underlying analytical methods themselves, rather than the evidence per se. We define this as a "methodology scoping review", which is becoming an essential component of research. Methodology scoping reviews provide a summary of existing analytical methods, techniques and software, proposed or applied in research articles, which address an analytical problem or further an analytical approach. These may be performed within a specific application domain (e.g. medicine) or may be entirely domain agnostic. These types of review can serve as a practical guide for applied researchers (by providing references to relevant articles, software, and tutorials), can identify methodological gaps that motivate new methods research or software, and can compare different approaches or solutions to an analytical problem in terms of their assumptions, applicability or computational complexity. Explicitly, the intention of methodology scoping reviews is usually to discover what methodology has been developed and/or applied to solve a particular analytical question. In some cases, one might wish to test and compare methods (e.g. through a simulation study [3,4]), but such comparison studies are usually performed after an initial methodology scoping review or methodology systematic review [5] to identify the relevant methods for comparison.

As an example of a methodology scoping review, our group recently synthesised the literature around analytical approaches to address calibration drift in clinical prediction models [6]. This review covered any analytical method that acknowledges the real-time collection of data and allowed regression coefficients of a prediction model to evolve over time [6]. Equally, there are many other examples of methodology scoping reviews within the literature (see, for example, [7–9]).

However, in contrast to systematic reviews, guidelines that promote best practice when conducting a methodology scoping review are limited. A recently published typology identified 14 different review types within the evidence synthesis literature [10]. Amongst these, the so-called 'scoping review' aims to consider a broader scope of literature than a systematic review, thereby providing comprehensive coverage of an area of literature, describing the volume of literature, and identifying gaps within current evidence bases [10–14]. Scoping reviews are perhaps the review type most closely aligned with the purpose of a methodology scoping review. They both attempt to summarise the state-of-play of a research area and provide a platform for future research, rather than for informing decisions directly. Importantly, scoping reviews have had clear guidelines published around their design and implementation [11–13] and are now considered a valid approach to evidence synthesis [15].

Despite these commonalities, in our experience there are several unique aspects to methodology scoping reviews that makes scoping review guidelines insufficient to adequately capture their design, implementation and reporting. Indeed, to the best of our knowledge, there are no guidelines on how to conduct reviews that aim to summarise analytical methods.

In this commentary article, we aim to discuss the potential of using existing scoping review guidelines in the context of methodology scoping reviews, and propose necessary extensions that draw on our recent experiences (e.g. [6,16,17]). We intend for this article to provide a foundational set of guidelines and motivate a wider discussion on the topic.

## Potential of Existing Review Strategies for Methods Research

Given the apparent overlaps between scoping reviews and methodology scoping reviews, we have used the existing scoping review guidelines as a foundation for our discussion of best practice in designing and conducting the latter. We refer readers to the wider literature for a detailed discussion [11–13], but in-short, a six-stage scoping review framework has been proposed that outlines how to: (1) identify the research question, (2) identify relevant studies, (3) select studies, (4) charter the data, (5) collate, summarise and report the findings, and (6) consult relevant stakeholders [11]. This framework was subsequently refined with additional recommendations on the six stages [12]. However, it is our experience that there are three areas that require additional consideration when specifically conducting a methodology scoping review, which we outline here.

### Identification of Search Terms

Current guidelines for scoping reviews state that the search terms should be defined to capture as broad a literature as possible. To a large extent, the same is true for methodology scoping reviews. Usually, the aim is to discover a broad range of analytical methods that have been – or could be – applied to solve a particular analytical/inference problem. For example, in the context of our review on dynamic prediction models, we aimed to understand the current state-of-the-art in dynamic prediction modelling and identify unsolved methodological challenges [6].

However, one challenge within methodology scoping reviews is that it is not normally possible to define the full list of search terms *a priori* since there is no equivalent of the Medical Subject Headings (MeSH) terminology index [18] for the methodological literature, and the language used for a particular method can vary across research domains [19]. For example, Gelman identified five different meanings for the distinction between fixed and random effects [20], while the (incorrect) interchangeable use of 'multivariate' and 'multivariable' is well-known [21]. Similarly, the same terminology can also be used to refer to different analytical problems or approaches. For example, one challenge in our dynamic modelling review for methods to address calibration drift [6], was that the term "dynamic model" is also used in survival analysis where one updates time-to-event predictions conditional on future survival [22]. This heterogeneity and homonymy in terminology across the analytical/statistical literature presents problems in defining the search terms and can lead to a large number of non-relevant literature within the search results.

**Identification of Methodological Papers: issues presented with the format of publishing**

Related to the challenge of identifying search terms, it can equally be challenging to identify papers that include potentially novel analytical methods, since the format of publishing often means that materials cannot easily be searched and screened. Specifically, methodological developments are not always presented in dedicated methods journals and therefore may be harder to identify.

For instance, new methods may be motivated by the peculiarities of a specific clinical question in a specific dataset that requires existing analytical approaches to be modified or extended. Here, the methodological novelty may be secondary to the clinical question and may therefore only be documented in brief or in supplementary materials, without a dedicated methods paper. Given the volume of literature in a given field, it is not usually viable to include all methods and applied papers, while simultaneously screening supplementary material. Therefore, some degree of selection is normally required, which increases the risk of missing important methodology.

Similarly, new analytical techniques may become available following developments or feature enhancements in statistical software, but these may be poorly documented and not published in the scientific literature. For example, advancements in the types of models a given software package can fit might only be highlighted through package release notes, rather than through a dedicated paper. The accessibility of such methods to applied researchers may therefore be difficult to determine.

**Scope and Content of the Review**

Given the heterogeneity in terminology and the breadth of methodology research, one should usually keep the methodology scoping review as broad as possible, but its breadth can result in a quantity of literature that is unfeasible to review completely given finite resource. This means that a level of pragmatism is needed to balance the breadth of included literature with time, budget and resource allocations.

Generally, scoping review guidance indicates that the scope can be modified iteratively so large volumes of papers can be screened quickly using abstracts [11–13]. Problematically, in many methodological papers, the new methodology is often not described in sufficient detail within abstracts. This makes it difficult to iteratively assess (using the abstracts alone) whether a paper should be included or excluded. When considering large volumes of abstracts against the inclusion/exclusion criteria of the methodology scoping review, this challenge can result in an increased number of false-positive papers that make it to full screening (since a conservative approach would dictate to full-screen any paper where it is uncertain, at abstract screening, if it meets the review inclusion criteria).

# Recommendations for Methodology scoping reviews

Based on the aforementioned challenges that we have faced when designing and conducting several methodology scoping reviews, we here suggest a set of foundational recommendations to supplement current scoping review guidelines [11–13]. These recommendations were formulated through a series of face-to-face workshops attended by the authors of this article, wherein personal experiences were shared and collated, to inform the below discussion (summarised in **Table 1**). Below, we structure these recommendations in a similar way to the scoping review guidelines [11–13].

**Identification of Review Scope**

As with all research studies, it is imperative to start the methodology scoping review with a clear understanding of what the research question is, to both inform the search strategy and limit the scope of the review. For example, within methodology scoping reviews, the research question could be to summarise the methodological field to serve as a practical guide for applied researchers. Alternatively, the purpose of the review could be to identify areas for further methodological development.

Once the reviewer establishes this aim, they can define the scope based on its intended use. For instance, if the aim of the review is to identify methodological gaps, then the reviewer should consider all research fields (e.g. not just medical literature). In contrast, a review that aims to provide a practical summary of methods for a given analytical problem, acting as a reference for applied researchers, could be limited to a particular application domain. Similar logic can be applied to the choice between inclusion of 'applied' articles (where the description of novel methods might be secondary to the research question) or dedicated methods papers. In other words, as with all evidence syntheses, the research question should inform the appropriate 'types' of papers that the reviewer wishes to include. For example, our recent review on dynamic modelling approaches aimed to identify methodological gaps and as such excluded applied research [6].

In all cases, we recommend that the context and scope of the methodology scoping review is articulated and reported clearly, to aid reproducibility and reporting standards.

**Identification of the Methodology Search Terms**

Given that it is unlikely that the reviewer will know all the terminology relating to the analytical methods *a-priori*, we have found that it is best to start by searching for a problem area, rather than specific methods. One approach is to start with a set of "key" papers that are known to the reviewer beforehand. If some of these key papers are previous review papers that explored similar topics, then one could utilise and modify the search terms therein. A second approach is where the reviewer might be aware of leading researchers that have worked on the indicated problem area. By using the keywords within the key papers or searching publications of leading researchers in a field, the reviewer can formulate an initial set of search terms. These initial search terms can then be used to find further papers, which again can inform a revised set of search terms. We refer to this iterative process as the Methodological Iterative Search Technique (MIST), which means that search terms can be expanded/collapsed as needed and as identified through resulting papers; this approach has been used successfully by some of the authors [6]. The MIST is similar to the Agile approach of systems design [23], and can help overcome some of the challenges associated with heterogeneity/ homonymy in language used for a given methodological area. Such iterative approaches have also been recommended elsewhere [24]. The MIST requires that the reviewer should remain open to changing the scope of the review based on the number of identified papers, again acknowledging that this needs to be done pragmatically to balance available resources. This process is similar, in principle, to how many healthcare reviews are conducted.

The MIST might not be relevant in all methodology scoping reviews (e.g. searching for methods with well-defined and standardised terminology, such as the missing data nomenclature). Moreover, its use could also reduce transparency and reproducibility, due to an inevitable level of subjectivity. As such, we propose that the MIST should be documented so that the reporting of the methodology scoping review clearly demonstrates how the search terms were modified over time. For example, this could be a flowchart that keeps track of the iterative changes to search terms, and is arguably the equivalent of a CONSORT diagram for methodology scoping reviews [25]. Moreover, registering a design/protocol of the methodology scoping review can aid the reproducibility of the MIST (see, for example, [16]). Transparent and reproducible reporting of the iterative approach can be achieved through repeated registrations of the review protocol (e.g., using the Open Science Framework), with appropriate version control. The timing of new registrations can be chosen to reflect key changes in the search strategy, scope or inclusion/exclusion criteria.

In addition, a useful feature of methodology scoping reviews is the so-called "snowballing" approach, whereby the reference list of screened papers is searched to identify additional relevant literature; likewise, one screens literature that has cited the included papers. This is a modification of the 'Pearl growing' technique, in the absence of MeSH terms [26,27]. Snowballing can aid the inclusion of papers from a range of disciplines and researchers, can help cover the breadth of literature, and can help capture any papers that were not picked-up by the search terms. Again, clearly documenting this process is needed to aid reproducibility.

**Selecting the Studies to Include**

In our experience, keeping a broad scope can increase confidence of finding relevant methodologies, but a degree of filtering is required to ensure the review remains feasible to complete. Namely, one could undertake a so-called "pre-review" between abstract and full-text review stages. Here, the reviewer should use pre-defined filters where possible, to ensure the number of papers for full-text review remains manageable. For example, there are now published filters for reviews that aim to find literature on clinical prediction models [28]. In the absence of such pre-defined filters, the reviewer could quickly (but systematically) pre-screen full texts to identify any papers that are 'related-to-the-review', but perhaps do not satisfy the full inclusion criteria. These papers might be worthy of mention in the methodology scoping review but will not likely constitute the main findings. We recommend that a second reviewer can help remove the potential subjectivity of this "pre-review" process.

**Collating, Summarising and Reporting the Findings**

Identifying the information to extract from identified papers can be a challenging part of conducting a methodology scoping review. The large variability in the way different analytical methods are developed, reported and implemented adds to this challenge. **Table 2** summarises data that one might wish to extract from a methodology scoping review.

We primarily recommend that the reviewer focusses on a description of the identified methods (i.e. what are the methods intended to be used for, and what analytical techniques underpin them?). Within a methodology scoping review, one does not usually test the methods against each other (in terms of analytical performance), but one might compare their characteristics (e.g. assumptions, applications, software, and limitations). Such comparisons will allow the reviewer to formulate themes and help with reporting the findings. For example, within our review on dynamic approaches to clinical prediction, we grouped all of the identified methodologies into three distinct themes based on their approaches to inference [6]. In saying this, we note that a large part of methodological research is indeed aiming to test and compare methods (in terms of analytical performance), perhaps through a simulation study (for example, [3,4]). In the context of how we define methodology scoping reviews, we propose that these are out of scope, but methodology scoping reviews could be used to inform the methods considered in such a methods comparison study.

Alternatively, software tutorials can be used to report the findings of methodology scoping reviews, which are worthy of specific mention. For example, many R packages have associated vignettes, where existing methodologies can be overviewed and summarised. This is a particularly useful way of describing the 'state-of-play' within a given methodological field and is ideal for those reviews that aim to provide practical guides for applied researchers.

## Conclusion

Conducting a methodology scoping review is similar, in principle, to a scoping review. As such, recent scoping review guidelines should be followed as much as possible [11–13]. However, it is our view and experience that these guidelines do not cover all aspects needed to sufficiently capture the design, implementation and reporting of methodology scoping reviews. To the best of our knowledge, such recommendations do not currently exist. Specifically, not all research-on-research studies (where the base unit of analysis is a controlled trial or observational study) are necessarily systematic reviews, and the use of systematic review guidelines in these cases may be inappropriate [29]; thus, new guidelines are warranted [30]. While our discussion presents a foundational set of recommendations, further research and discussion is needed in this space. An agreed set of guidelines surrounding methodology scoping reviews is key to ensuring they are designed, conducted and reported in a standardised and reproducible manner. This is vital given the importance of methodology scoping reviews in informing and directing methodological research.

## Declarations

**Competing Interests**

The authors declare that they have no competing interests

**Funding**

None

**Authors' Contributions**

GPM came up with the initial idea for the paper and drafted the initial version of the manuscript. GPM, DAJ, LB, RS, LL, WH, AW, WW, MB, CSP and AP took part in a series of face-to-face workshops to discuss and develop the ideas we present in this paper. All authors (GPM, DAJ, LB, RS, LL, WH, AW, WW, MB, CSP, AP, MS and NP) contributed to the development and writing of the paper and approved the final version.

**Acknowledgements**

The Predictive Healthcare Analytics Group comprises the following people: Glen Martin, Niels Peek, Matthew Sperrin, David Jenkins, Anthony Wilson, Lucy Bull, Rose Sisk, Lijing Lin, William Hulme, Wenjuan Wang, Michael Barrowman, Camilla Sammut-Powell and Alexander Pate.

# Tables

**Table 1:** Summary of recommendations for the design and implementation for methodology scoping reviews, mapped to existing scoping review guidelines.

| Stage | Recommendation on top of current scoping review guidance |
|---|---|
| 1. Identification of Research Question | Clearly define whether the aim of the review is to summarise previous methodological research for the purposes of identifying gaps (e.g. for further methodological development) or if the purpose is to summarise methodological literature for applied researchers. The research question should be presented in a non-contextualised format. |
| 2. Identification of Relevant Studies | Take a pragmatic and iterative approach to identify search terms. Start by identifying "key" search terms either from papers known to the reviewers *a priori* or by searching papers from known leading researchers in the field. This initial set of search terms can then be refined iteratively to achieve a broad set of search terms to cover the literature. Clear documentation of this process is key to aid reproducibility by allowing others to see how search terms were modified over time in this iterative approach. We refer to this iterative process as the Methodological Iterative Search Technique (MIST). |
| 3. Study Selection | A choice usually needs to be made between reviewing papers where methods have been developed/applied to specific domains, or if the review seeks to summarise methodological advancement irrespective of application. Equally, a choice should be made regarding whether the review will only consider literature in the primary domain of interest, or if it will also include other domains. |
| 4. Charting the Data | We recommend that the reviewer focusses on a description of the identified methods (i.e. what are the methods intended to be used for, and what analytical techniques underpin them?). Additionally, the assumptions, inference approach, limitations and applied domains. Further information can be found in Table 2. |
| 5. Collating, summarising and reporting the results | The way in which the identified methodologies are summarised and reported will largely depend on the results of each review. In general, we recommend aiming to group the identified methodologies into distinct themes based on their assumptions, inference/ estimation approaches or application. |
| 6. Consultation | We recommend following the same guidelines for consultation as previously published for scoping reviews [11–13]. |

**Table 2**: Potential features to extract from identified papers within a methodology scoping review

| Feature | Motivation |
|---|---|
| Name of the methodology/ overview of the approach to inference | Often new analytical methods have a 'name' associated with them to allow other researchers to refer to them. Extracting such information from papers (or at least understanding the inferential techniques underpinning them) can aid in developing the set of themes. |
| Reported advantages/disadvantages of the method | All analytical methods have assumptions and/or limitations, which could restrict the domains in which they can be reasonably be implemented. Extracting this information can help to identify a practical guide of when each method can/should be used, and also identify areas of future methodological research. |
| Reported similarities with other methods | Sometimes, novel analytical methods are an extension of existing methods, or several methods might share common analytical/ inference properties. Extracting such information can aid in grouping a range of analytical methods into related themes. |
| Software to implement the methods | This is particularly important if the intended aim of the methodology review is to provide a practical guide for other researchers wishing to apply methodology. Collating available R/SAS/Stata/Python packages can help in this regard. |
| Field/discipline (e.g. journal) | Summarising the field of literature that new analytical methods have been developed and/or implemented in can aid in understanding what additional research (if any) is needed for them to be applied in a specific domain of interest. It can also help identify areas that might benefit from the methods. |
| Manuscript information (e.g. year of publication, or number of citations) | This information is primarily useful to summarise the uptake of a given method within the applied literature (e.g. changes over time). |

| Indication of whether the paper/study was applied (and if so, what was the application/domain?), or purely methodological (and if so, was there a motivating example?) | Reporting on the range of areas that a given method has been developed and/or applied in can help in identifying if the methods are applied to data of a particular format/structure. |
|---|---|